\documentclass[%
 aip,
 sd,%
 amsmath,amssymb,
 reprint,%
]{revtex4-1}

\usepackage{graphicx}% Include figure files
\usepackage{dcolumn}% Align table columns on decimal point
\usepackage{bm}% bold math
\usepackage{amsmath}
\usepackage[english]{babel}

\usepackage{siunitx}

\begin{document}

\preprint{AIP/123-QED}

\title[Surface acoustic wave devices on bulk ZnO crystals at low temperature]{Surface acoustic wave devices on bulk ZnO crystals at low temperature}% Force line breaks with \\
%\thanks{Footnote to title of article.}

\author{E.\,B. Magnusson}
%\email{einar.magnusson@physics.ox.ac.uk}
\author{B.\,H. Williams}

\author{R. Manenti}

\author{M.-S. Nam}

\author{A. Nersisyan}
\author{M.\,J. Peterer}

\author{A. Ardavan}%
\author{P.\,J. Leek}
\affiliation{ 
Clarendon Laboratory, Department of Physics, University of Oxford, Parks Road, Oxford OX1 3PU, United Kingdom
%Authors' institution and/or address%\\This line break forced with \textbackslash\textbackslash
}%

\date{\today}% It is always \today, today,
             %  but any date may be explicitly specified

\begin{abstract}
Surface acoustic wave (SAW) devices based on thin films of ZnO are a well established technology. However, SAW devices on bulk ZnO crystals are not practical at room temperature due to the significant damping caused by finite electrical conductivity of the crystal. Here, by operating at low temperatures, we demonstrate effective SAW devices on the (0001) surface of bulk ZnO crystals, including a delay line operating at SAW wavelengths of $\lambda=4$ and $6$\,{\si\um} and a one-port resonator at a wavelength of $\lambda= 1.6$ \si\um. We find that the SAW velocity is temperature dependent, reaching $v\simeq2.68$ km/s at 10\,mK. Our resonator reaches a maximum quality factor of $Q_i\simeq1.5 \times 10^5$, demonstrating that bulk ZnO is highly viable for low temperature SAW applications. The performance of the devices is strongly correlated with the bulk conductivity, which quenches SAW transmission above about 200\,K.
\end{abstract}

%\pacs{Valid PACS appear here}% PACS, the Physics and Astronomy
                             % Classification Scheme.
%Use showkeys class option if keyword
                              %display desired
\maketitle

\section{Introduction}

Surface acoustic waves (SAWs) are mechanical wave modes confined to the surface of a material. In piezoelectric materials, a SAW is accompanied by an electric field and can thus be used for implementing electrical signal processing devices such as filters and oscillators, widely used in telecommunications \cite{Morgan2007}. The response of a SAW device can also be very sensitive to external parameters such as temperature, pressure or mass loading of the surface, making them excellent candidates for a variety of sensors including biosensors, in which a functionalised surface binds to specific molecules \cite{Lange2008}.

SAWs are usually excited by means of a periodic AC electric field, requiring a piezoelectric material to be present to couple electric field to mechanical stress. Two configurations are commonly deployed: either the material (bulk) substrate is piezoelectric, or a thin layer of a piezoelectric material is deposited onto a non-piezoelectric substrate.
To excite a SAW, an oscillating voltage of frequency $f_\text{SAW}$ is applied to an interdigital transducer \cite{White1965} (IDT) whose spacing determines the wavelength $\lambda_0$, such that $f_\text{SAW} \lambda_0 = v_\text{SAW}$. The SAW velocity $v_\text{SAW}$ is in the range $1 - 7 \text{ km/s}$ for commonly used single-crystal materials \cite{Morgan2007} but can reach $12 \text{ km/s}$ for layered systems \cite{Nakahata2003, Lin2007}. The IDT also works as a receiver, generating an oscillating voltage when a SAW of the right wavelength passes through it \cite{Morgan2007}.
A SAW reflector similar to an optical Bragg mirror may be implemented by a metal grating or grooves in the surface.
Using these components, it is possible to construct delay lines (by spatially separating transmitting and receiving IDTs) and resonators (by enclosing an IDT between two reflectors \cite{Morgan2007, Bell1976}).

ZnO is a relatively common material in SAW devices, but until now 
there have only been reports of its use as a thin layer piezoelectric transducer on top of a non-piezoelectric substrate such as sapphire, diamond or SiO$_2$/Si\cite{Morgan2007, Weber}. 
Wafer scale bulk ZnO has recently become available\cite{Look1998, Look2001, Avrutin2010}, but even nominally pure material contains dopants that give rise to a non-zero room-temperature electrical conductivity. This conductivity damps SAWs efficiently, impeding the use of bulk ZnO in SAW devices.

Due in part to the development of bulk crystal growth, ZnO is receiving growing interest as a device material \cite{Avrutin2010, Jagadish2006, Ozgur2005}.
Its electronic bandstructure, in particular its wide bandgap and large exciton binding energy, make it promising for optoelectronic applications such as ultra-violet light, and exciton based emitters \cite{Ozgur2005}.
ZnO is also less susceptible to radiation damage than similar materials, making it a good candidate for space applications \cite{Look2001}.  ZnO has also recently been investigated as a possible material to be used in quantum information devices, due to the presence of long lifetime spin defect centres in the crystal \cite{George2013}.

In this paper we report measurements of SAW devices on the (0001) plane of high quality bulk wurtzite ZnO \cite{MTI}. We work at reduced temperatures to freeze out itinerant charge carriers, thereby decreasing the bulk electrical conductivity and the resultant SAW damping. We fabricate the devices using electron beam lithography and liftoff of a 5/50nm Ti/Al evaporated bilayer. We cool the devices down to a lowest temperature of 10 mK using a dilution refrigerator, and measure their frequency response using a vector network analyzer (VNA). We also perform two-contact electrical resistance measurements as a function of temperature on samples cut from the same ZnO wafer, using graphite paste for contacts, and a liquid $^4$He dewar dipstick fitted with a Cernox thermometer.

\section{Measurements}

We measure delay lines with a short effective path length to measure temperature dependence at high temperatures, where we expect losses to be high. To observe low-temperature effects, we measure an under-coupled high quality resonator that is much more sensitive to any dissipation due to its long effective path length.

\subsection{Delay line}

\begin{figure}[t]
\includegraphics[width=0.99\linewidth]{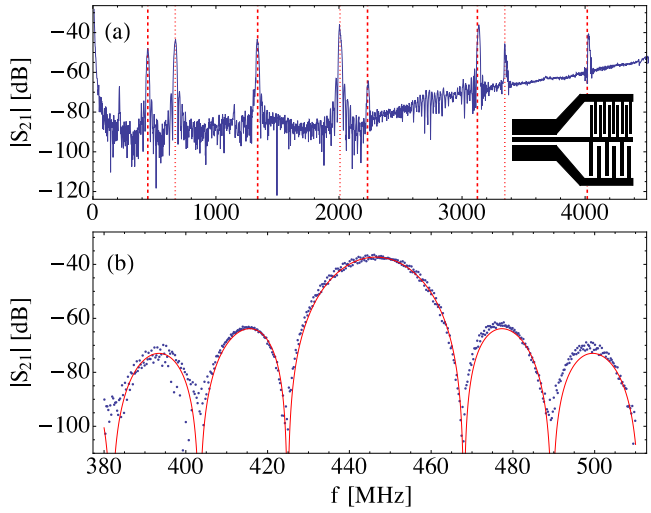}
\caption{\label{fig:delayline} 
(a) Transmission spectrum of a delay line with $4$ {\si\um} and $6$ {\si\um} transducers. Multiple odd harmonics are seen. The $6$ {\si\um} fundamental mode at $f=446$ MHz and its third, fifth, seventh and ninth harmonics are marked with vertical dashed lines, and the $6$ {\si\um} modes up to fifth harmonic are marked with dotted lines. (b) Zoom of transmission spectrum at the fundamental mode ($f=446$ MHz) of the delay line at 50 mK. The solid line is a fit of Eq.\,\ref{IDT_S21_Eq} to the data. Inset: Schematic of dual wavelength IDT connected to coplanar waveguide (not to scale).}
\end{figure}

The delay line device has two input transducers in parallel, on either side of the centre conductor of a coplanar waveguide input line (see inset of Fig. \ref{fig:delayline}(b)). One operates at 
a SAW wavelength of $\lambda=4$ {\si\um} and the other at $\lambda=6$ \si\um. Each IDT has 20 finger pairs, and at
2\,mm separation, there is an identical mirrored transducer for output. This parallel IDT design is used simply for convenience of gathering more data with a single device. 

Fig.\,\ref{fig:delayline}(a) shows the transmission of the delay line, $S_{21}$, as a function of frequency at 50\,mK. The attenuation of the cables in the cryostat has been subtracted. Both the $\lambda=6$\,{\si\um} and $\lambda=4$\,{\si\um} transducers are active.
We observe multiple transmission peaks, 
with the fundamentals at $f_1=446.4$ MHz (arising from the $\lambda=6$ {\si\um} transducers) and $f_2=669.5$ MHz ($\lambda=4$ {\si\um} transducers), 
implying a SAW velocity of $v \simeq 2678$ m/s under the IDTs.
We find higher harmonics in the transmission spectrum at $3 f_1$, $3 f_2$, $5 f_1$, $5 f_2$, $7 f_1$ and $9 f_1$, demonstrating the possibility to operate bulk ZnO SAW devices up to at least $4~{\rm GHz}$. 

Simple interdigital transducers of the kind we use here are expected to give rise to a transmission spectrum close to the fundamental frequency that depends on frequency as 
\begin{equation}
|S_{21}| = A \,\text{sinc}^4\left(  N \Delta f /f_0 \right) 
\label{IDT_S21_Eq}
\end{equation} 
where $N$ is the number of IDT fingers, and $\Delta f$ is the detuning from the center frequency $f_0$\,\cite{Morgan2007}. Fig.\,\ref{fig:delayline}(b) shows a measurement of the transmission spectrum close to $f_1$ and a fit to Eq.\,\ref{IDT_S21_Eq}, demonstrating that the behaviour of our device is close to ideal.

\begin{figure}[b]
\includegraphics[width=0.99\linewidth]{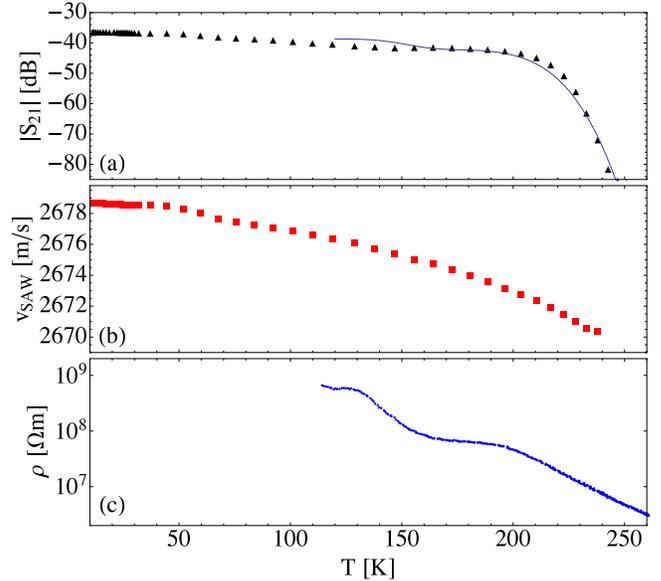} 
\caption{\label{fig:delaylinetemp} 
(a) Transmission amplitude $S_{21}$ of the fundamental mode ($f=446 \text{ MHz}$) of the delay line as a function of temperature $T$. The solid line is a fit of transmission attenuating linearly with conductivity. (b) SAW velocity as a function of temperature, calculated by fitting the peak frequency and using $v_{\text SAW}=\lambda_0/f_{\text SAW}$. (c) Resistivity of the ZnO substrate as a function of temperature. }
\end{figure}

Fig.\,\ref{fig:delaylinetemp}(a) shows the transmission of the delay line measured at $f_1$ as a function of temperature. The temperature dependent attenuation of the cryostat cables has been calibrated out. The delay line transmission is only weakly temperature dependent below 50\,K, but drops slowly at higher temperatures up to about 200\,K, above which it drops very sharply. This decrease in transmission is accompanied by a shift in $f_1$; fitting the transmission to Eq.\,\ref{IDT_S21_Eq} to obtain $f_1$ at each temperature allows us to derive the change in the SAW velocity with temperature, as shown in Fig.\,\ref{fig:delaylinetemp}(b). 

At low temperatures, the ZnO substrate is not electrically conductive. However, at elevated temperatures thermal excitation of impurities leads to $n$-type doping and a non-zero conductivity\,\cite{Ozgur2005}. Fig.\,\ref{fig:delaylinetemp}(c) shows the resistivity of our ZnO substrate as a function of temperature. Below about 120\,K, the resistance is too high to measure using our apparatus. At higher temperatures, there are several intervals in temperature where the resistivity drops, which may be due to different impurities becoming ionised. These changes in resistivity are correlated with drops in $|S_{21}|$, because the electric field component of the SAW can drive dissipative currents, thereby damping the SAW. The solid line is a fit of the function $f(T) = A \, e^{-B/\rho(T)}$ to the data, where $A = -38.3\text{ dB}$ and $B = 2.65 \times 10^8 ~\Omega $m are fit parameters. This model assumes that SAW dissipation is linear in the DC conductivity of the ZnO. Deviations from this simple model may be due to the frequency dependent absorption of different defect centres in the crystal.

\subsection{Resonator}

The resonator has a single IDT with 21 fingers ($\lambda=1.6 \text{ \si\um}$) and is measured in reflection. On either side
of the transducer there are passive gratings with 1750 fingers, separated from the IDT by the width of a finger, meaning that there is no discontinuity between the IDT and grating. 

\begin{figure}[t]
\includegraphics[width=0.99\linewidth]{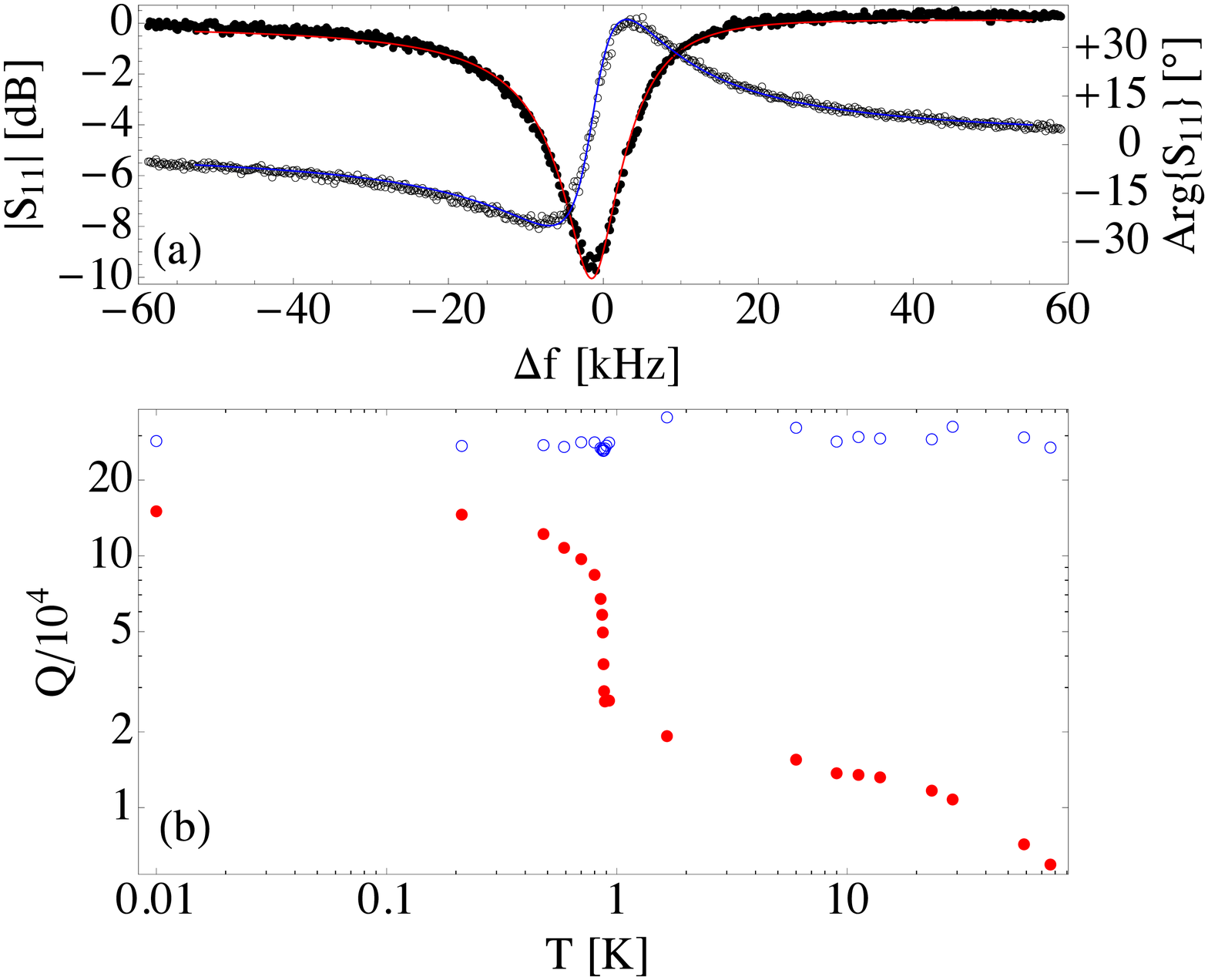} 
\caption{\label{fig:resonator} 
(a) Reflection S11 of the one-port SAW resonator as a function of frequency deviation from resonance $\Delta f$, showing amplitude $|S_\text{11}|$ (phase $\text{Arg}\{S_\text{11}\}$) in filled (open) circles. The full complex reflection $S_\text{11}$ is fitted to Eq.\,\ref{refl_coeff} (solid lines).
(b) Dependence of resonator internal quality factor $Q_i$ (filled circles) and external quality factor $Q_e$ (empty circles) on temperature. }
\end{figure}

Fig.\,\ref{fig:resonator}(a) shows the amplitude and phase of the reflection $S_{11}$ from the one-port resonator as a function of frequency around its resonant frequency of $f_0\simeq1.677784 \text{ GHz}$, measured at an input power of $-50 \text{ dBm}$ and at $T=10 \text{ mK}$. In Fig.\,\ref{fig:resonator}(a), the reflection is normalised to its value far from resonance, and a linear phase shift with frequency (arising from the finite path length of the measurement cables) has been removed. Multiple additional resonances were also observed at nearby frequencies, corresponding to other harmonics of the resonator that lie within the IDT bandwidth (not shown). 

As a function of the deviation $\Delta f$ of the frequency from resonance, the complex reflection coefficient is given by\,\cite{Manenti2014} 
\begin{equation}
S_{11} 
=1-\frac{2Q_i/Q_e}{(Q_e+Q_i)/Q_e+i2Q_i \Delta f/f_0}e^{i\phi_0}
\label{refl_coeff}
\end{equation}
where $Q_e$ is the external quality factor (due to the input port coupling), 
$Q_i$ is the internal quality factor (intrinsic to the resonator), $\Delta f= (f-f_0)$ is the detuning from the resonance frequency $f_0$, and $\phi_0$ is a parameter accounting for the slight asymmetry of the resonance due to impedance mismatches \cite{Megrant2012}. 
The complex data is fitted to Eq.\,\ref{refl_coeff} and for the data in Fig.\,\ref{fig:resonator}(a), the fit yields $f_0 = 1.677784 \text{ GHz}$, $Q_e = 2.78\times 10^5$ and $Q_i = 1.49\times 10^5$. 

Fig.\,\ref{fig:resonator}(b) shows the dependence of $Q_i$ (filled circles) and $Q_e$ (empty circles) on temperature. As the temperature increases from its base value, the quality factor $Q_i$ decreases, and is particularly strongly temperature dependent as $T$ approaches $1~\rm{K}$. This dependence is likely to be attributable to quasiparticle scattering in the Ti/Al bilayer, and its transition from the superconducting to normal state at $T \simeq 850\rm{mK}$.
In contrast, $Q_e$ is observed to be approximately temperature independent. Aside from the fixed IDT geometry, $Q_e$ is dependent on several properties of the substrate; the dielectric and piezoelectric coupling constants, and $v_\text{SAW}$ \cite{Morgan2007}. We can conclude therefore that these properties are not strongly temperature-dependent in this range.

% dielectric constant
% piezo coupling coefficient

At base temperature, the quality factor $Q_i$ reaches values near $1.5 \times 10^5 $, which is close to the highest quality factors reported in SAW resonators fabricated on other materials \cite{ElHabti1996}. It is interesting to note that in our device, superconducting electrodes increase $Q_i$ by about a factor of 5.

\section{Conclusions}

We have reported SAW devices on a bulk ZnO crystal substrate. Our results show that bulk ZnO devices are highly feasible at low temperatures, with resonator quality factors among the highest reported, while at high temperature, losses correlate with the bulk DC electrical conductivity. Such high quality SAW devices may find use in low temperature physics research such as quantum information science, where SAWs have already started playing a role\,\cite{Gustafsson2012,Gustafsson2014}. 

\section{Acknowledgements}

We would like to thank A.\,Patterson for technical contributions to the project, and A.\,Landig for measurements of superconducting thin films. This work has received funding from the UK Engineering and Physical Sciences Research Council under grants EP/J001821/1, EP/J013501/1 and EP/G003610/1, and the People Programme (Marie Curie Actions) of the European Union's Seventh Framework Programme (FP7/2007-2013) under REA grant agreement n$^\text{o}$ [304029]. EBM acknowledges the Clarendon Fund and the Jesus College Old MembersÕ Award for financial support.

\bibliography{ZnO_SAW}% Produces the bibliography via BibTeX.

%merlin.mbs aipnum4-1.bst 2010-07-25 4.21a (PWD, AO, DPC) hacked
%Control: key (0)
%Control: author (8) initials jnrlst
%Control: editor formatted (1) identically to author
%Control: production of article title (0) allowed
%Control: page (1) range
%Control: year (1) truncated
%Control: production of eprint (0) enabled
\begin{thebibliography}{19}%
\makeatletter
\providecommand \@ifxundefined [1]{%
 \@ifx{#1\undefined}
}%
\providecommand \@ifnum [1]{%
 \ifnum #1\expandafter \@firstoftwo
 \else \expandafter \@secondoftwo
 \fi
}%
\providecommand \@ifx [1]{%
 \ifx #1\expandafter \@firstoftwo
 \else \expandafter \@secondoftwo
 \fi
}%
\providecommand \natexlab [1]{#1}%
\providecommand \enquote  [1]{``#1''}%
\providecommand \bibnamefont  [1]{#1}%
\providecommand \bibfnamefont [1]{#1}%
\providecommand \citenamefont [1]{#1}%
\providecommand \href@noop [0]{\@secondoftwo}%
\providecommand \href [0]{\begingroup \@sanitize@url \@href}%
\providecommand \@href[1]{\@@startlink{#1}\@@href}%
\providecommand \@@href[1]{\endgroup#1\@@endlink}%
\providecommand \@sanitize@url [0]{\catcode `\\12\catcode `\$12\catcode
  `\&12\catcode `\#12\catcode `\^12\catcode `\_12\catcode `\%12\relax}%
\providecommand \@@startlink[1]{}%
\providecommand \@@endlink[0]{}%
\providecommand \url  [0]{\begingroup\@sanitize@url \@url }%
\providecommand \@url [1]{\endgroup\@href {#1}{\urlprefix }}%
\providecommand \urlprefix  [0]{URL }%
\providecommand \Eprint [0]{\href }%
\providecommand \doibase [0]{http://dx.doi.org/}%
\providecommand \selectlanguage [0]{\@gobble}%
\providecommand \bibinfo  [0]{\@secondoftwo}%
\providecommand \bibfield  [0]{\@secondoftwo}%
\providecommand \translation [1]{[#1]}%
\providecommand \BibitemOpen [0]{}%
\providecommand \bibitemStop [0]{}%
\providecommand \bibitemNoStop [0]{.\EOS\space}%
\providecommand \EOS [0]{\spacefactor3000\relax}%
\providecommand \BibitemShut  [1]{\csname bibitem#1\endcsname}%
\let\auto@bib@innerbib\@empty
%</preamble>
\bibitem [{\citenamefont {Morgan}(2007)}]{Morgan2007}%
  \BibitemOpen
  \bibfield  {author} {\bibinfo {author} {\bibfnamefont {D.}~\bibnamefont
  {Morgan}},\ }\href
  {http://www.amazon.com/Surface-Acoustic-Filters-Second-Edition/dp/0123725372}
  {\emph {\bibinfo {title} {{Surface Acoustic Wave Filters, Second Edition:
  With Applications to Electronic Communications and Signal Processing (Studies
  in Electrical and Electronic Engineering)}}}}\ (\bibinfo  {publisher}
  {Academic Press},\ \bibinfo {year} {2007})\BibitemShut {NoStop}%
\bibitem [{\citenamefont {L\"{a}nge}, \citenamefont {Rapp},\ and\ \citenamefont
  {Rapp}(2008)}]{Lange2008}%
  \BibitemOpen
  \bibfield  {author} {\bibinfo {author} {\bibfnamefont {K.}~\bibnamefont
  {L\"{a}nge}}, \bibinfo {author} {\bibfnamefont {B.}~\bibnamefont {Rapp}}, \
  and\ \bibinfo {author} {\bibfnamefont {M.}~\bibnamefont {Rapp}},\ }\bibfield
  {title} {\enquote {\bibinfo {title} {{Surface acoustic wave biosensors: a
  review}},}\ }\href {\doibase 10.1007/s00216-008-1911-5} {\bibfield  {journal}
  {\bibinfo  {journal} {Analytical and Bioanalytical Chemistry}\ }\textbf
  {\bibinfo {volume} {391}},\ \bibinfo {pages} {1509--1519} (\bibinfo {year}
  {2008})}\BibitemShut {NoStop}%
\bibitem [{\citenamefont {White}\ and\ \citenamefont
  {Voltmer}(1965)}]{White1965}%
  \BibitemOpen
  \bibfield  {author} {\bibinfo {author} {\bibfnamefont {R.~M.}\ \bibnamefont
  {White}}\ and\ \bibinfo {author} {\bibfnamefont {F.~W.}\ \bibnamefont
  {Voltmer}},\ }\bibfield  {title} {\enquote {\bibinfo {title} {{DIRECT
  PIEZOELECTRIC COUPLING TO SURFACE ELASTIC WAVES}},}\ }\href@noop {}
  {\bibfield  {journal} {\bibinfo  {journal} {Applied Physics Letters}\
  }\textbf {\bibinfo {volume} {7}} (\bibinfo {year} {1965})}\BibitemShut
  {NoStop}%
\bibitem [{\citenamefont {Nakahata}\ \emph {et~al.}(2003)\citenamefont
  {Nakahata}, \citenamefont {Fujii}, \citenamefont {Higaki}, \citenamefont
  {Hachigo}, \citenamefont {Kitabayashi}, \citenamefont {Shikata},\ and\
  \citenamefont {Fujimori}}]{Nakahata2003}%
  \BibitemOpen
  \bibfield  {author} {\bibinfo {author} {\bibfnamefont {H.}~\bibnamefont
  {Nakahata}}, \bibinfo {author} {\bibfnamefont {S.}~\bibnamefont {Fujii}},
  \bibinfo {author} {\bibfnamefont {K.}~\bibnamefont {Higaki}}, \bibinfo
  {author} {\bibfnamefont {A.}~\bibnamefont {Hachigo}}, \bibinfo {author}
  {\bibfnamefont {H.}~\bibnamefont {Kitabayashi}}, \bibinfo {author}
  {\bibfnamefont {S.}~\bibnamefont {Shikata}}, \ and\ \bibinfo {author}
  {\bibfnamefont {N.}~\bibnamefont {Fujimori}},\ }\bibfield  {title} {\enquote
  {\bibinfo {title} {{Diamond-based surface acoustic wave devices}},}\ }\href
  {\doibase 10.1088/0268-1242/18/3/314} {\bibfield  {journal} {\bibinfo
  {journal} {Semiconductor Science and Technology}\ }\textbf {\bibinfo {volume}
  {18}},\ \bibinfo {pages} {S96--S104} (\bibinfo {year} {2003})}\BibitemShut
  {NoStop}%
\bibitem [{\citenamefont {Lin}\ \emph {et~al.}(2007)\citenamefont {Lin},
  \citenamefont {Wu}, \citenamefont {Chen},\ and\ \citenamefont
  {Chou}}]{Lin2007}%
  \BibitemOpen
  \bibfield  {author} {\bibinfo {author} {\bibfnamefont {C.-M.}\ \bibnamefont
  {Lin}}, \bibinfo {author} {\bibfnamefont {T.-T.}\ \bibnamefont {Wu}},
  \bibinfo {author} {\bibfnamefont {Y.-Y.}\ \bibnamefont {Chen}}, \ and\
  \bibinfo {author} {\bibfnamefont {T.-T.}\ \bibnamefont {Chou}},\ }\bibfield
  {title} {\enquote {\bibinfo {title} {{Improved frequency responses of SAW
  filters with interdigitated interdigital transducers on ZnO/Diamond/Si
  layered structure}},}\ }\href@noop {} {\bibfield  {journal} {\bibinfo
  {journal} {J. Mech.}\ }\textbf {\bibinfo {volume} {23}},\ \bibinfo {pages}
  {253} (\bibinfo {year} {2007})}\BibitemShut {NoStop}%
\bibitem [{\citenamefont {{Bell D.L.T.}}\ and\ \citenamefont
  {Li}(1976)}]{Bell1976}%
  \BibitemOpen
  \bibfield  {author} {\bibinfo {author} {\bibfnamefont {J.}~\bibnamefont
  {{Bell D.L.T.}}}\ and\ \bibinfo {author} {\bibfnamefont {R.~C.~M.}\
  \bibnamefont {Li}},\ }\bibfield  {title} {\enquote {\bibinfo {title}
  {{Surface-acoustic-wave resonators}},}\ }\href {\doibase
  10.1109/PROC.1976.10200} {\bibfield  {journal} {\bibinfo  {journal}
  {Proceedings of the IEEE}\ }\textbf {\bibinfo {volume} {64}},\ \bibinfo
  {pages} {711--721} (\bibinfo {year} {1976})}\BibitemShut {NoStop}%
\bibitem [{\citenamefont {Weber}, \citenamefont {Weiss},\ and\ \citenamefont
  {Hunklinger}(1991)}]{Weber}%
  \BibitemOpen
  \bibfield  {author} {\bibinfo {author} {\bibfnamefont {A.}~\bibnamefont
  {Weber}}, \bibinfo {author} {\bibfnamefont {G.}~\bibnamefont {Weiss}}, \ and\
  \bibinfo {author} {\bibfnamefont {S.}~\bibnamefont {Hunklinger}},\ }\bibfield
   {title} {\enquote {\bibinfo {title} {{Comparison of Rayleigh and Sezawa wave
  modes in ZnO-SiO\_2-Si structures}},}\ }in\ \href {\doibase
  10.1109/ULTSYM.1991.234187} {\emph {\bibinfo {booktitle} {IEEE 1991
  Ultrasonics Symposium}}}\ (\bibinfo  {publisher} {IEEE},\ \bibinfo {year}
  {1991})\ pp.\ \bibinfo {pages} {363--366}\BibitemShut {NoStop}%
\bibitem [{\citenamefont {Look}\ \emph {et~al.}(1998)\citenamefont {Look},
  \citenamefont {Reynolds}, \citenamefont {Sizelove}, \citenamefont {Jones},
  \citenamefont {Litton}, \citenamefont {Cantwell},\ and\ \citenamefont
  {Harsch}}]{Look1998}%
  \BibitemOpen
  \bibfield  {author} {\bibinfo {author} {\bibfnamefont {D.}~\bibnamefont
  {Look}}, \bibinfo {author} {\bibfnamefont {D.}~\bibnamefont {Reynolds}},
  \bibinfo {author} {\bibfnamefont {J.}~\bibnamefont {Sizelove}}, \bibinfo
  {author} {\bibfnamefont {R.}~\bibnamefont {Jones}}, \bibinfo {author}
  {\bibfnamefont {C.}~\bibnamefont {Litton}}, \bibinfo {author} {\bibfnamefont
  {G.}~\bibnamefont {Cantwell}}, \ and\ \bibinfo {author} {\bibfnamefont
  {W.}~\bibnamefont {Harsch}},\ }\bibfield  {title} {\enquote {\bibinfo {title}
  {{Electrical properties of bulk ZnO}},}\ }\href {\doibase
  10.1016/S0038-1098(97)10145-4} {\bibfield  {journal} {\bibinfo  {journal}
  {Solid State Communications}\ }\textbf {\bibinfo {volume} {105}},\ \bibinfo
  {pages} {399--401} (\bibinfo {year} {1998})}\BibitemShut {NoStop}%
\bibitem [{\citenamefont {Look}(2001)}]{Look2001}%
  \BibitemOpen
  \bibfield  {author} {\bibinfo {author} {\bibfnamefont {D.}~\bibnamefont
  {Look}},\ }\bibfield  {title} {\enquote {\bibinfo {title} {{Recent advances
  in ZnO materials and devices}},}\ }\href {\doibase
  10.1016/S0921-5107(00)00604-8} {\bibfield  {journal} {\bibinfo  {journal}
  {Materials Science and Engineering: B}\ }\textbf {\bibinfo {volume} {80}},\
  \bibinfo {pages} {383--387} (\bibinfo {year} {2001})}\BibitemShut {NoStop}%
\bibitem [{\citenamefont {Avrutin}\ \emph {et~al.}(2010)\citenamefont
  {Avrutin}, \citenamefont {Cantwell}, \citenamefont {Song}, \citenamefont
  {Silversmith},\ and\ \citenamefont {Morko\c{c}}}]{Avrutin2010}%
  \BibitemOpen
  \bibfield  {author} {\bibinfo {author} {\bibfnamefont {V.}~\bibnamefont
  {Avrutin}}, \bibinfo {author} {\bibfnamefont {G.}~\bibnamefont {Cantwell}},
  \bibinfo {author} {\bibfnamefont {J.~J.}\ \bibnamefont {Song}}, \bibinfo
  {author} {\bibfnamefont {D.~J.}\ \bibnamefont {Silversmith}}, \ and\ \bibinfo
  {author} {\bibfnamefont {H.}~\bibnamefont {Morko\c{c}}},\ }\bibfield  {title}
  {\enquote {\bibinfo {title} {{Bulk ZnO: Current Status, Challenges, and
  Prospects}},}\ }\href {\doibase 10.1109/JPROC.2010.2040363} {\bibfield
  {journal} {\bibinfo  {journal} {Proceedings of the IEEE}\ }\textbf {\bibinfo
  {volume} {98}},\ \bibinfo {pages} {1339--1350} (\bibinfo {year}
  {2010})}\BibitemShut {NoStop}%
\bibitem [{\citenamefont {Jagadish}\ and\ \citenamefont
  {Pearton}(2006)}]{Jagadish2006}%
  \BibitemOpen
  \bibinfo {editor} {\bibfnamefont {C.}~\bibnamefont {Jagadish}}\ and\ \bibinfo
  {editor} {\bibfnamefont {S.}~\bibnamefont {Pearton}},\ eds.,\ \href
  {http://www.sciencedirect.com/science/book/9780080447223} {\emph {\bibinfo
  {title} {{Zinc Oxide Bulk, Thin Films and Nanostructures}}}},\ \bibinfo
  {edition} {1st}\ ed.\ (\bibinfo  {publisher} {Elsevier},\ \bibinfo {address}
  {Oxford},\ \bibinfo {year} {2006})\BibitemShut {NoStop}%
\bibitem [{\citenamefont {Özgür}\ \emph {et~al.}(2005)\citenamefont
  {Özgür}, \citenamefont {Alivov}, \citenamefont {Liu}, \citenamefont
  {Teke}, \citenamefont {Reshchikov}, \citenamefont {Doğan}, \citenamefont
  {Avrutin}, \citenamefont {Cho},\ and\ \citenamefont {Morkoç}}]{Ozgur2005}%
  \BibitemOpen
  \bibfield  {author} {\bibinfo {author} {\bibfnamefont {U.}~\bibnamefont
  {Özgür}}, \bibinfo {author} {\bibfnamefont {Y.~I.}\ \bibnamefont
  {Alivov}}, \bibinfo {author} {\bibfnamefont {C.}~\bibnamefont {Liu}},
  \bibinfo {author} {\bibfnamefont {A.}~\bibnamefont {Teke}}, \bibinfo {author}
  {\bibfnamefont {M.~A.}\ \bibnamefont {Reshchikov}}, \bibinfo {author}
  {\bibfnamefont {S.}~\bibnamefont {Doğan}}, \bibinfo {author} {\bibfnamefont
  {V.}~\bibnamefont {Avrutin}}, \bibinfo {author} {\bibfnamefont {S.-J.}\
  \bibnamefont {Cho}}, \ and\ \bibinfo {author} {\bibfnamefont
  {H.}~\bibnamefont {Morkoç}},\ }\bibfield  {title} {\enquote {\bibinfo
  {title} {{A comprehensive review of ZnO materials and devices}},}\ }\href
  {\doibase 10.1063/1.1992666} {\bibfield  {journal} {\bibinfo  {journal}
  {Journal of Applied Physics}\ }\textbf {\bibinfo {volume} {98}},\ \bibinfo
  {pages} {041301} (\bibinfo {year} {2005})}\BibitemShut {NoStop}%
\bibitem [{\citenamefont {George}, \citenamefont {Edwards},\ and\ \citenamefont
  {Ardavan}(2013)}]{George2013}%
  \BibitemOpen
  \bibfield  {author} {\bibinfo {author} {\bibfnamefont {R.~E.}\ \bibnamefont
  {George}}, \bibinfo {author} {\bibfnamefont {J.~P.}\ \bibnamefont {Edwards}},
  \ and\ \bibinfo {author} {\bibfnamefont {A.}~\bibnamefont {Ardavan}},\
  }\bibfield  {title} {\enquote {\bibinfo {title} {{Coherent Spin Control by
  Electrical Manipulation of the Magnetic Anisotropy}},}\ }\href {\doibase
  10.1103/PhysRevLett.110.027601} {\bibfield  {journal} {\bibinfo  {journal}
  {Physical Review Letters}\ }\textbf {\bibinfo {volume} {110}},\ \bibinfo
  {pages} {027601} (\bibinfo {year} {2013})}\BibitemShut {NoStop}%
\bibitem [{MTI()}]{MTI}%
  \BibitemOpen
  \href@noop {} {}\bibinfo {note} {We studied nominally pure single crystal
  samples obtained commercially from the MTI Corporation,
  http://mtixtl.com.}\BibitemShut {Stop}%
\bibitem [{\citenamefont {Manenti}(2014)}]{Manenti2014}%
  \BibitemOpen
  \bibfield  {author} {\bibinfo {author} {\bibfnamefont {R.}~\bibnamefont
  {Manenti}},\ }\emph {\bibinfo {title} {{Surface Acoustic Waves for Quantum
  Information}}},\ \href@noop {} {Ph.D. thesis},\ \bibinfo  {school}
  {University of Oxford} (\bibinfo {year} {2014})\BibitemShut {NoStop}%
\bibitem [{\citenamefont {Megrant}\ \emph {et~al.}(2012)\citenamefont
  {Megrant}, \citenamefont {Neill}, \citenamefont {Barends}, \citenamefont
  {Chiaro}, \citenamefont {Chen}, \citenamefont {Feigl}, \citenamefont {Kelly},
  \citenamefont {Lucero}, \citenamefont {Mariantoni}, \citenamefont
  {O’Malley}, \citenamefont {Sank}, \citenamefont {Vainsencher},
  \citenamefont {Wenner}, \citenamefont {White}, \citenamefont {Yin},
  \citenamefont {Zhao}, \citenamefont {Palmstro̸m}, \citenamefont {Martinis},\
  and\ \citenamefont {Cleland}}]{Megrant2012}%
  \BibitemOpen
  \bibfield  {author} {\bibinfo {author} {\bibfnamefont {A.}~\bibnamefont
  {Megrant}}, \bibinfo {author} {\bibfnamefont {C.}~\bibnamefont {Neill}},
  \bibinfo {author} {\bibfnamefont {R.}~\bibnamefont {Barends}}, \bibinfo
  {author} {\bibfnamefont {B.}~\bibnamefont {Chiaro}}, \bibinfo {author}
  {\bibfnamefont {Y.}~\bibnamefont {Chen}}, \bibinfo {author} {\bibfnamefont
  {L.}~\bibnamefont {Feigl}}, \bibinfo {author} {\bibfnamefont
  {J.}~\bibnamefont {Kelly}}, \bibinfo {author} {\bibfnamefont
  {E.}~\bibnamefont {Lucero}}, \bibinfo {author} {\bibfnamefont
  {M.}~\bibnamefont {Mariantoni}}, \bibinfo {author} {\bibfnamefont {P.~J.~J.}\
  \bibnamefont {O’Malley}}, \bibinfo {author} {\bibfnamefont
  {D.}~\bibnamefont {Sank}}, \bibinfo {author} {\bibfnamefont {A.}~\bibnamefont
  {Vainsencher}}, \bibinfo {author} {\bibfnamefont {J.}~\bibnamefont {Wenner}},
  \bibinfo {author} {\bibfnamefont {T.~C.}\ \bibnamefont {White}}, \bibinfo
  {author} {\bibfnamefont {Y.}~\bibnamefont {Yin}}, \bibinfo {author}
  {\bibfnamefont {J.}~\bibnamefont {Zhao}}, \bibinfo {author} {\bibfnamefont
  {C.~J.}\ \bibnamefont {Palmstro̸m}}, \bibinfo {author} {\bibfnamefont
  {J.~M.}\ \bibnamefont {Martinis}}, \ and\ \bibinfo {author} {\bibfnamefont
  {A.~N.}\ \bibnamefont {Cleland}},\ }\bibfield  {title} {\enquote {\bibinfo
  {title} {{Planar superconducting resonators with internal quality factors
  above one million}},}\ }\href {\doibase 10.1063/1.3693409} {\bibfield
  {journal} {\bibinfo  {journal} {Applied Physics Letters}\ }\textbf {\bibinfo
  {volume} {100}},\ \bibinfo {pages} {113510} (\bibinfo {year}
  {2012})}\BibitemShut {NoStop}%
\bibitem [{\citenamefont {{El Habti}}(1996)}]{ElHabti1996}%
  \BibitemOpen
  \bibfield  {author} {\bibinfo {author} {\bibfnamefont {A.}~\bibnamefont {{El
  Habti}}},\ }\bibfield  {title} {\enquote {\bibinfo {title} {{High-frequency
  surface acoustic wave devices at very low temperature: Application to loss
  mechanisms evaluation}},}\ }\href {\doibase 10.1121/1.415953} {\bibfield
  {journal} {\bibinfo  {journal} {The Journal of the Acoustical Society of
  America}\ }\textbf {\bibinfo {volume} {100}},\ \bibinfo {pages} {272}
  (\bibinfo {year} {1996})}\BibitemShut {NoStop}%
\bibitem [{\citenamefont {Gustafsson}\ \emph {et~al.}(2012)\citenamefont
  {Gustafsson}, \citenamefont {Santos}, \citenamefont {Johansson},\ and\
  \citenamefont {Delsing}}]{Gustafsson2012}%
  \BibitemOpen
  \bibfield  {author} {\bibinfo {author} {\bibfnamefont {M.~V.}\ \bibnamefont
  {Gustafsson}}, \bibinfo {author} {\bibfnamefont {P.~V.}\ \bibnamefont
  {Santos}}, \bibinfo {author} {\bibfnamefont {G.}~\bibnamefont {Johansson}}, \
  and\ \bibinfo {author} {\bibfnamefont {P.}~\bibnamefont {Delsing}},\
  }\bibfield  {title} {\enquote {\bibinfo {title} {{Local probing of
  propagating acoustic waves in a gigahertz echo chamber}},}\ }\href {\doibase
  10.1038/nphys2217} {\bibfield  {journal} {\bibinfo  {journal} {Nature
  Physics}\ }\textbf {\bibinfo {volume} {8}},\ \bibinfo {pages} {338--343}
  (\bibinfo {year} {2012})}\BibitemShut {NoStop}%
\bibitem [{\citenamefont {Gustafsson}\ \emph {et~al.}(2014)\citenamefont
  {Gustafsson}, \citenamefont {Aref}, \citenamefont {Kockum}, \citenamefont
  {Ekstr\"{o}m}, \citenamefont {Johansson},\ and\ \citenamefont
  {Delsing}}]{Gustafsson2014}%
  \BibitemOpen
  \bibfield  {author} {\bibinfo {author} {\bibfnamefont {M.~V.}\ \bibnamefont
  {Gustafsson}}, \bibinfo {author} {\bibfnamefont {T.}~\bibnamefont {Aref}},
  \bibinfo {author} {\bibfnamefont {A.~F.}\ \bibnamefont {Kockum}}, \bibinfo
  {author} {\bibfnamefont {M.~K.}\ \bibnamefont {Ekstr\"{o}m}}, \bibinfo
  {author} {\bibfnamefont {G.}~\bibnamefont {Johansson}}, \ and\ \bibinfo
  {author} {\bibfnamefont {P.}~\bibnamefont {Delsing}},\ }\bibfield  {title}
  {\enquote {\bibinfo {title} {{Propagating phonons coupled to an artificial
  atom.}}}\ }\href {\doibase 10.1126/science.1257219} {\bibfield  {journal}
  {\bibinfo  {journal} {Science (New York, N.Y.)}\ }\textbf {\bibinfo {volume}
  {346}},\ \bibinfo {pages} {207--11} (\bibinfo {year} {2014})}\BibitemShut
  {NoStop}%
\end{thebibliography}%

\end{document}